\documentstyle[preprint,floats,aps,pre]{revtex}
\begin{document}
\draft
\title{Two-Well System under Large Amplitude Periodic Forcing:
Stochastic Synchronization, Stochastic Resonance and Stability}

\author{Mangal C. Mahato and A.M. Jayannavar}
\address{Institute of Physics, Sachivalaya Marg,
Bhubaneswar-751005, India}

\maketitle

\begin{abstract}
We study the residence time distributions and explore the
possibility of observing stochastic resonance and synchonization
of passages in a two-well system driven by a periodic forcing of
amplitude larger than a marginal value beyond which one of the
two wells become unstable and diasppear. We define and calculate
hysteresis loop in the system, the area of which measures the
degree of synchronization between the residence time statistics
and the input signal, as a function of input noise strength. We
analyse the noise induced stability obtained in such a
deterministically overall unstable system and within this
context discuss the above two phenomena. 
\end{abstract}


Stochastic resonance (SR) is a nonlinear phenomena. It helps a
system respond to an input signal with an enhanced output
signal, as reflected in the power at the same frequency of its
spectrum by partially rectifying the noise concomitant to the
input signal. This
phenomena\cite{Benzi,McNamara1,Jung1,Gamma1,Fox,Nato} is
observed experimentally\cite{Fauve,McNamara2,Douglass} as well
as in numerical\cite{Benzi,McNamara1,Casado,MCM} and analog
simulations\cite{Gamma2,Debnath} as a peak when the output
signal to (background) noise ratio (SNR) in the power spectrum
is plotted as a function of input noise strength. This could
find practical application in detecting weak signals and
understandably theoretical treatments and numerical simulations,
except in a few cases\cite{Dayan,Jung2}, are mostly focussed on
small amplitude periodic input signals. In this work we study a
two-well system with large (overcritical) amplitude periodic
forcing (input signal) aided by zero-mean Gaussian white noise
and explore the possibility and physical origin of observing SR
in the distribution of residence times of a Brownian particle in
any one of the two wells.  SR is observed in a well defined
range of sweep rates of the input signal. We find that SR in
this case is made possible by noise-affected slowing down of the
passages of the particle from an unstable well to the stable
well of the two-well system.  Also, this slowing down has marked
effect on the synchronization of passages. This will be
discussed below after describing the procedure of our
calculation.

We represent the two-well system by the usual Landau potential
\begin{equation}
U(m)=-\frac{a}{2}m^2+\frac{b}{4}m^4
\end{equation}
and solve the overdamped Langevin equation
\begin{equation}
\dot m=-\frac{\partial U(m)}{\partial m}+h(t)+\hat f(t),
\end{equation}
for the motion $m(t)$, when the system is subjected to a time
dependent external forcing $h(t)$ and a Gaussian random
fluctuating force $\hat f(t)$ with statistics $<\hat f(t)>=0$
and $<\hat f(t)\hat f(t')>=2D\delta (t-t')$. We take $a=2.0$ and
$b=1.0$ throughout our calculation and $h(t)$ to be a saw-tooth
type external periodic field with amplitude $h_0=1.4h_c$, where
$\vert h_c\vert$ is the minimum value of field at which one of
the two wells of $\Phi (m)=U(m)-mh(t)$ become unstable and then
diassapear. We set $m(t=0)$ in one of the two well minima (the
initial condition is not really important) and monitor the time
evolution $m(t)$ for a long time (corresponding to a large
number of passages) to obtain meaningful average values. If the
particle at some instance $t$ is in a well, we say the passage
to take place at a later time to the other well only when the
trajectory $m(t)$ traverses across the inflection point on to
the other side of the potential barrier separating the first
well from the other. We put markers on the time axis whenever a
passage takes place, the interval between two consecutive
markers giving the residence time in a well. We thus obtain the
residence time distributions $\rho_1(\tau)$ and $\rho_2(\tau)$
in the two wells respectively. (In our present case the two
distributions should be identical.) The markers on the time axis
also give the field values $h(t)$ at which passage takes place
and thus the jump field distributions $\rho _{12}(h)$ and $\rho
_{21}(h)$ are obtained for passages from well 1 to 2 , and from
well 2 to 1, respectively. These distributions, $\rho _i(t)$ and
$\rho _{ij}(h)$ are used below in our discussion of SR (by
calculating the SNR from the power spectrum\cite{Pres} of $\rho
_i$'s) and stochastic synchronization (SS)(by calculating the
hysteresis loop area) in the two-well system at large forcing
amplitude $h_0=1.4h_c$. As $h(t)$ varies periodically, the left
well, for example, becomes unstable and disappears for the part
of the period for which $h(t)\geq h_c$ and so is the other well
for the part of the period for which $h(t)\leq -h_c$. During
these two intervals (due to the nonexistence of the intervening
potential barrier) passage to the stable well is possible even
when $D=0$. For small sweep rates $\vert \dot h \vert$, of
$h(t)$, one passage from one well to the other per period is
guaranteed for $D=0$. However, for large $\vert \dot h \vert$
the passage may be delayed and the distribution $\rho _i(t)$ may
spill over to the next and even later periods of $h(t)$. The
multipeaked $\rho _i(t)$ indicates the possibility of SR at
large sweep rates as $D$ is gradually increased. And, indeed, we
find the system to show stochastic resonant behaviour.

The exact value of $D$ at which the stochastic resonance occurs
is difficult to pinpoint as the uncertainties involved in
measuring the output noise level is large. However, the
occurrence of resonant behaviour can be seen clearly from Fig.
1, where we plot the power spectral density (of $\rho (t)$) for
various $D$ values: Almost overlapping broad peakes with
appreciable height appear only for a small intermediate range of
$D$ values. One may, however, ask whether this is not an
artefact of having large sweep rates and, thus, having no
relation with real physical situations. This doubt can be set to
rest once we look into an another important question of noise
induced stability of "unstable states".

As stated earlier, for small enough $\vert\dot h\vert$, the
average number of passages per cycles (ANPPC) of $h(t)$ at $D=0$
is 1 but starts decreasing sharply to zero as $\vert \dot
h\vert$ is increased beyond a threshold value $\dot h_{th}$
(in the present case $\approx 0.384h_c$) as shown in Fig. 2.
Hence till $\vert\dot h\vert= \dot h_{th}$ the sweep rate is
such that the particle gets enough time to roll down to the
stable well in each and every cycle in the absence of friction
(noise).Now, let us look at the variation of ANPPC of $h(t)$ as
a function of $D$ for various $\vert \dot h\vert$ [Fig. 3]. The
curves for $\vert \dot h\vert<\dot h_{th}$ show that as the
noise is switched on ANPPC decreases to less than 1, attains a
minimum and then rises to 1 and beyond 1 corresponding to
processes dominated by noise. This is an important observation
and shows clearly that the presence of noise actually slows down
the decay of the deterministically overall unstable states in an
appropriate range of $\vert \dot h \vert$ and $D$ values. This
fact has been reported earlier\cite{Dayan,Mantegna} too in other
systems. In order to make our discussion on SR sensible we need,
therefore, to restrict $\vert \dot h \vert$ to the maximum value
of $\dot h_{th}$. Fig. 1 corresponds to $\dot h =0.28h_c$ which
is well within the range, and the observation of SR in the case
of large amplitude periodic forcings can, thus, safely be
considered to be genuine and not an artefact of large sweep
rates. Having discussed about the SR behaviour reflected in the
residence time distribution we ask the question of
synchronization property of the passages with reference to the
input signal (periodic forcing). 

The phase relationship of passages with the input signal is well
reflected in the distributions $\rho_{ij}(h)$ of field values
$h(t)$ at which passages take place from well $i$ to well $j$.
In order to obtain a quantitative relationship we invoke the
hysteretic property of the system. The hysteresis loop is
defined here as the difference, $\bar m(h)$, of relative
populations $m_1(h)$ and $m_2(h)$ in the two wells 1 and 2,
respectively. Here $m_2(h)$ is calculated from the discrete
equation, 
\begin{equation}
m_2(h)=m_2(h-\Delta h)-m_2(h-\Delta h) \rho_{21}(h)\Delta h
+m_1(h-\Delta h)\rho_{12}(h)\Delta h,
\end{equation}
with negligibly small $\Delta h$, and similarly for
$m_1(h)=1-m_2(h)$ and with the condition of closure of the loop
in a cycle of $h(t)$. The area of the hysteresis loop is a good
measure of the degree of synchronization of passages. For
instance, if the passages from well 1 to the well 2 take place
only at $h=h_0$, where the barrier to passage $1\rightarrow 2$
is minimum, and similarly from well 2 to well 1 only at $h=-h_0$
(a case of maximum synchronization) then the hysteresis loop is
a rectangle with the largest area, whereas if the passages take
place all over randomly and uniformly from $h_0$ to $-h_0$, we
have the least area, equal to zero, a case of least
synchronization between passages and the input periodic signal.
On plotting the hysteresis loop area as a function of $D$ for
various $\vert \dot h\vert$ (Fig. 4), we find that the area
peaks in the range of $D$ where we find noise induced delay of
passages. However, these peak positions, in general, do not
coincide with the peak positions of SNR. This stochastic
synchronization of passages may not, however, be entirely
unrelated to stochastic resonance\cite{MCMJ}. 

In conclusion, we observe that SR is a meaningful phenomena at
large amplitude periodic forcings too. It is because of the
presence of noise that the particle stochastically fails, in
some of the field sweep cycles, to roll down to the stable well
so that the residence time distribution contains more than one
peak. For a lone peak in the residence time distribution {\it
may not} yield SR. Noise of appropriate strength prolongs the
decay of the deterministically overall unstable state and
thereby it helps to maximize the degree of synchronization of
passages (with respect to the externally applied periodic field)
in the two-well system.

\vfil
\newpage
\begin{figure}
\caption{Shows power spectral density (psd) for three
values of $D$: (a) $D=.001$ (dotted), (b) $D=.1$(solid line), and
(c) $D=.4$ (dash-dotted), at $\dot h=.028h_c$. We have taken, for
each curve, 256 (augmented with zeroes) data points at a time
interval of 2.0. }
~~~~~

\caption{Average number of passages per cycle (ANPPC) is
plotted as a function of $\dot h$ for $D=0.00001$. Averaging is
done over 5000 cycles. The line is drawn to guide the eye.}
~~~~~

\caption{Shows ANPPC as a function of $D$ for various
$\dot h$ values:(a) $0.05h_c(\circ)$, (b) $0.2h_c (\Box)$, (c)
$0.28h_c (\diamond)$, (d) $0.35h_c (\triangle)$, and (e) $0.4h_c
(\triangleleft)$. For $\dot h\widetilde{>}0.392h_c$ the curves,
for example the curve for $\dot h=.4h_c$, show monotonic
behaviour starting from ANPPC=0. The lines are drawn to guide
the eye.} 
~~~~~

\caption{Hysteresis loop area $A$ versus $D$ for various
$\dot h$ values:(a) $0.05h_c (\circ)$, (b) $0.2h_c (\Box)$, (c)
$0.28h_c (\diamond)$, (d) $0.35h_c (\triangle)$, and (e) $0.4h_c
 (\triangleleft)$. The lines are drawn to guide the eye.} 
\end{figure}
~~~~~~

~~~~~
\end{document}